\documentclass[9pt,twocolumn,twoside]{opticajnl}
\journal{opticajournal} 

\setboolean{shortarticle}{true}

\usepackage{braket}
\usepackage{lineno}
\usepackage{physics}

\title{Desynchronization of temporal solitons in Kerr cavities with pulsed injection}

\author[1*]{D. A. Dolinina}
\author[2]{G. Huyet}
\author[3]{D. Turaev}
\author[1]{A. G. Vladimirov}

\affil[1]{Weierstrass Institute, Mohrenstr. 39, 10117 Berlin, Germany}
\affil[2]{Universit\'e C\^ote d’Azur, CNRS, INPHYNI, 06200 Nice, France}
\affil[3]{Imperial College, London SW7 2AZ, United Kingdom}
\affil[*]{dolinina@wias-berlin.de}

\begin{abstract}
A numerical and analytical study was conducted to investigate the bifurcation mechanisms that cause desynchronization between the
soliton repetition frequency and the frequency of external pulsed injection in a Kerr cavity described by the Lugiato-Lefever equation. The results suggest that desynchronization typically occurs through an Andronov-Hopf bifurcation. Additionally, a simple and intuitive criterion for this bifurcation to occur is proposed. 
\end{abstract}

\setboolean{displaycopyright}{false} 

\begin{document}

\maketitle

Optical frequency combs have had a significant impact on several fields including spectroscopy, optical ranging, metrology, exoplanet search, microwave photonics and optical communications \cite{cundiff2003colloquium,schroder2019laser,picque2019frequency,suh2019searching,trocha2018ultrafast,xue2016microwave}. 
A conventional approach to frequency comb generation involves the use of optical microresonators. In particular, considerable attention has been paid to microcavity soliton frequency combs, which have been experimentally observed in \cite{Herr14}. These combs are characterised by the generation of temporal cavity solitons (TCSs), which are stable, periodic light pulses that maintain their shape as they propagate. In simpler setups, TCSs are generated by injecting a continuous wave (CW) laser into a microcavity. However, the use of pulsed injection can be advantageous as it allows a reduction in the TCS excitation energy, a potential improvement in their properties, and the ability to tune the TCS repetition frequency by synchronising it with the injection pulse repetition frequency.
On the other hand, to achieve this synchronization, the repetition frequency of the injected pulses must be close to or a multiple of the free spectral range of the cavity.
Therefore, it is important to study the locking range and understand how it depends on the microcavity and external injection parameters.

A standard theoretical tool for describing TCS formation in microcavities is the paradigmatic Lugiato-Lefever equation (LLE) \cite{lugiato1987spatial}. The standard LLE is unable to describe the overlap of resonances corresponding to different cavity modes, unlike  the infinite-dimensional Ikeda map model \cite{blow1984global,hansson2015frequency}, the locally injected LLE \cite{kartashov2017multistability,conforti2017multi}, and the neutral delay differential equation (DDE) Kerr cavity model \cite{vladimirov2023neutral}. Nevertheless it has proven to be a very efficient tool for describing high-finesse microcavities used for optical frequency comb generation. 
The formation of 1D dissipative solitons in the LLE under constant injection is well studied, see e.g. \cite{barashenkov1996existence, barashenkov1998bifurcation,godey2014stability,parra2018bifurcation}. Theoretical studies of microcavity TCS generation by slowly modulated and pulsed injection have been carried out using the LLE in \citep{xu2014experimental,luo2015resonant,malinowski2017optical,li2022efficiency,cardoso2017localized,wang2018addressing,hendry2018spontaneous,hendry2019impact,hendry2020novel,erkintalo2022phase,talenti2023control}. 
For slightly modulated injection, an equation governing the slow time evolution of the TCS coordinate has been derived using an asymptotic approach \cite{erkintalo2022phase}. This equation is applicable to describe the dynamics of the TCS in the case of a pulsed pump source, where the injection pulse width is much larger than that of the TCS \cite{hendry2018spontaneous,hendry2019impact,hendry2020novel,talenti2023control}. The asymptotic equation for TCS motion in the presence of a small frequency mismatch between the injection pulse repetition rate and the cavity free spectral range (FSR), leading to TCS drift, was investigated in \cite{hendry2018spontaneous,hendry2019impact,hendry2020novel,erkintalo2022phase,talenti2023control}. It was shown that even with zero frequency mismatch, a symmetry-breaking bifurcation can occur as the pulse peak power increases, resulting in a shift of the TCS position from the peak of the injection pulse to its periphery. 

Here, using the LLE, we comprehensively investigate the bifurcation mechanisms leading to the unlocking between the repetition rates of the injection pulse and the TCS in a synchronously pumped optical microcavity. We show that for a sufficiently broad injection pulse, unlocking occurs via an Andronov-Hopf (AH) bifurcation rather than the saddle-node (SN) bifurcation responsible for the disappearance of the stationary TCS, as predicted by the TCS drift equation. Furthermore, we introduce a simple asymptotic criterion for the occurrence of the AH bifurcation, which requires only the knowledge of the injection pulse shape and the TCS solution with homogeneous injection. This semi-analytical criterion shows excellent agreement with results derived from numerical simulations of the LLE.

The paradigmatic Lugiato-Lefever equation (LLE) \cite{lugiato1987spatial} is a widely used tool for studying the dynamics of the electromagnetic field in Kerr resonators with coherent external injection, especially in microcavities used for optical frequency comb generation \cite{maleki2010high,matsko2011mode,lugiato2018lugiato}. This equation can be derived from the Maxwell-Bloch equations under the slowly varying envelope approximation \cite{lugiato2018lugiato}. Recently, it has been shown that the LLE can be obtained using a multiscale approach of Ref.~\cite{kolokolnikov2006q} from the neutral delay differential equation model \cite{vladimirov2023neutral} of an externally injected ring Kerr cavity. The resulting LLE, neglecting third and higher order dispersion terms, can be expressed in the following form
\begin{equation}
    \dfrac{\partial A}{\partial t}= -V \dfrac{\partial A}{\partial \xi} + i \dfrac{\partial^2 A}{\partial \xi ^2} + i |A|^2 A  -\left(1 + i\theta\right) A + \eta(\xi) .
    \label{eq:PDE_V}
\end{equation}
In this context, $A(\xi,t)$ is the normalized electric field envelope, $t$ is the ``slow'' time, $\xi$ is the ``fast'' time, and $\theta$ is the normalized detuning of the pump laser from the nearest cavity resonant frequency. The Kerr nonlinearity coefficient, the second-order dispersion coefficient, and the cavity decay rate are normalized to unity by rescaling the field amplitude, the ``fast'' time, and the ``slow'' time variable, respectively. The drift parameter $V$ defines the small frequency difference between the repetition rate of the input laser pulses and the free-spectral range of the cavity. The parameter $\eta$ represents the external coherent injection. In numerical calculations, we use Gaussian shape of the injection pulses 
$\eta(\xi) = p_0 \exp\left[-(d + i c)\xi^2\right]$,
where $p_0$ is the amplitude of the pulse, 
The parameter $d$ ($c$) determines width (chirp) of the injection pulse.   



\begin{figure}[ht]
\centering
\includegraphics[width=0.95\linewidth]{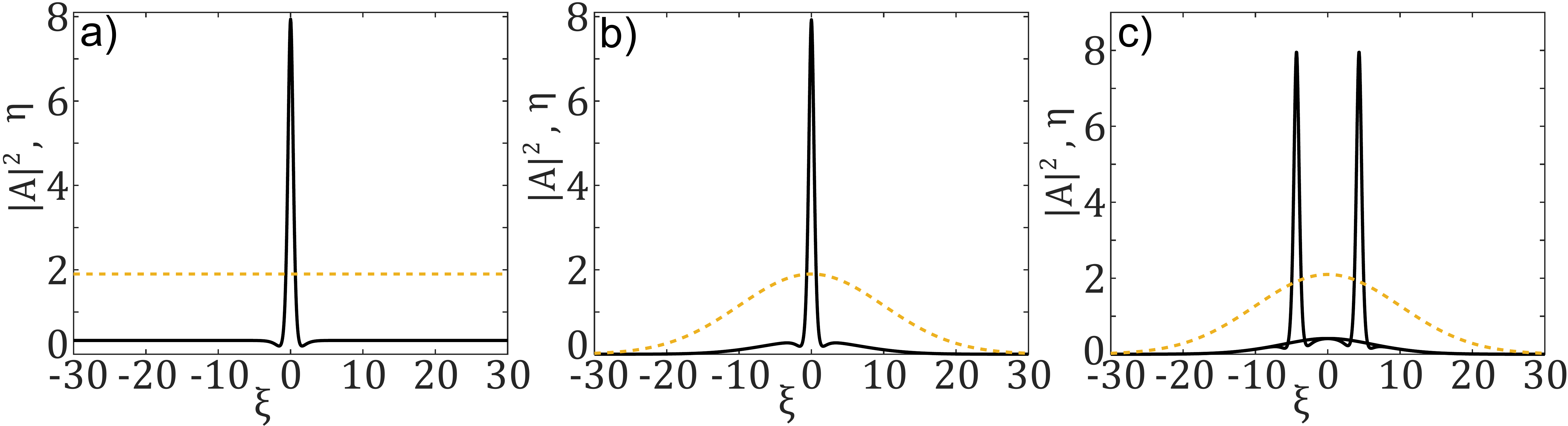}
\caption{Stationary TCS of Eq.~(\ref{eq:PDE_V}) under  homogeneous $\eta = 1.9$ (a) and Gaussian pulse injection with  $p_0 = 1.9<p_c$  (b) and  $p_0 = 2.1>p_c$ (c). For all panels: $\theta = 3.5$, $V=0$, $d = 0.005$ and $c=0$.   Yellow dashed lines show the injection distribution. Black curves show TCS intensities. Here and everywhere in calculations the length of the system is $L = 100$.}
\label{fig:sols}
\end{figure}

Stationary TCS solutions computed numerically with homogeneous injection and relatively wide Gaussian injection pulses with $V = 0$ (zero detuning between cavity FSR and pulse repetition rate) are shown in Fig.~\ref{fig:sols}. It can be seen that wide pulse injection has little effect on the shape of the TCS compared to the homogeneous case. As shown in previous studies \cite{hendry2018spontaneous,hendry2020novel}, the position of the stationary TCS depends on the amplitude of the injection pulse. When the amplitude is relatively small, the TCS rests at the center of the pulse where the injection is at its maximum, see Fig.~\ref{fig:sols}(b). When the injection amplitude exceeds a critical value, $p_0>p_c$, a spontaneous symmetry-breaking bifurcation occurs, causing the TCS to lose stability at the center of the pulse. As a result of this pitchfork bifurcation, two stable solutions appear whose stationary positions are shifted in both directions towards the periphery of the injection pulse, see Fig.~\ref{fig:sols}(c). 
Such a difference in TCS stability is caused by a drift due to the inhomogeneity of the injection, which pushes the TCS towards or away from the peak of the pulse, depending on the injection amplitude.

\begin{figure}[ht]
\centering
\includegraphics[width=0.9\linewidth]{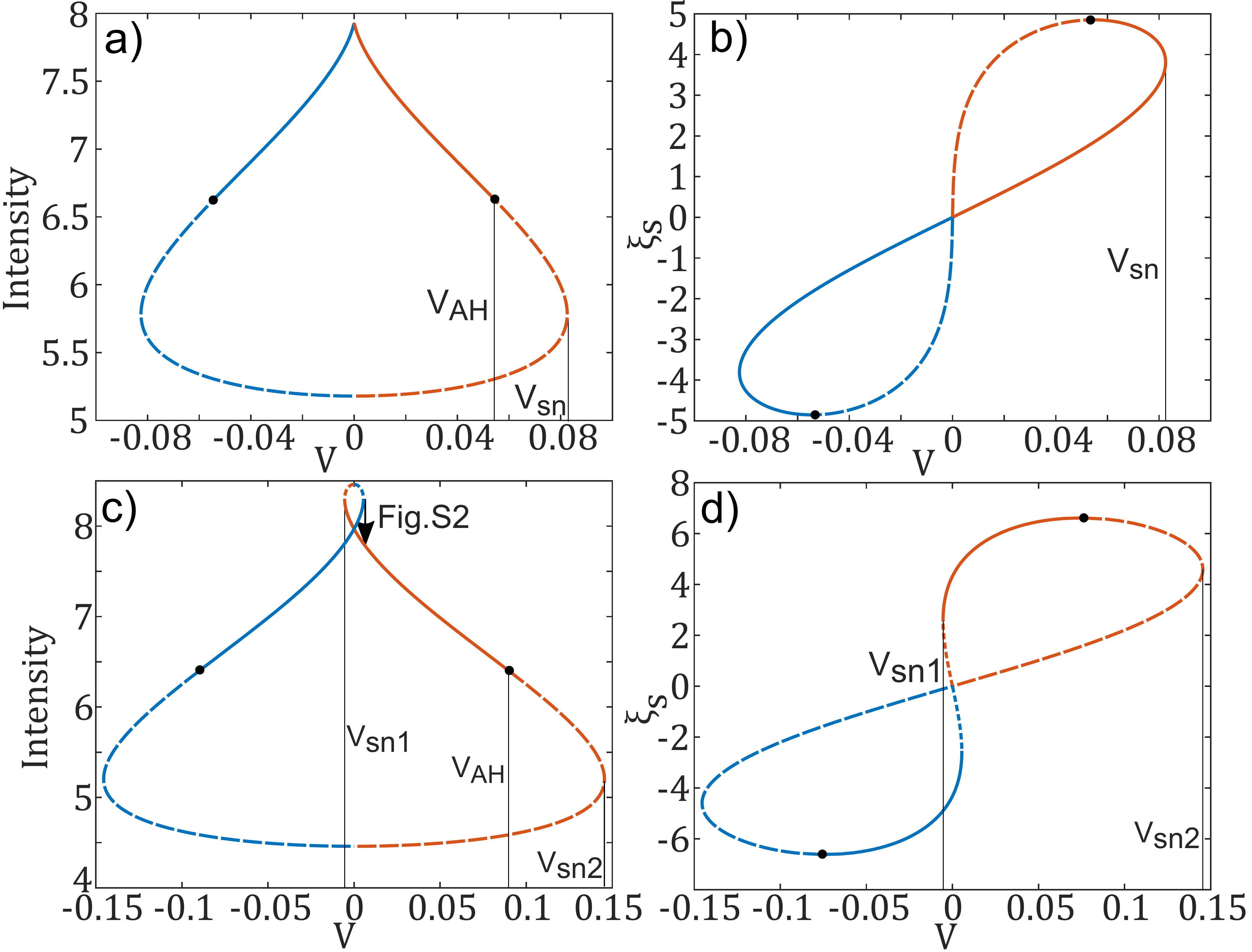}
\caption{TCS peak intensity (a,c) and position $\xi_s$ (b,d) as functions of $V$.  Upper (lower) panels correspond to $p_0 = 1.9<p_c$ ($p_0 = 2.1>p_c$). Stable and unstable solutions are indicated by solid and dashed lines, respectively. Other parameters are $\theta = 3.5$, $d = 0.005$, $c = 0$.}
\label{fig:synch_reg}
\end{figure}

\begin{figure}[ht]
\centering
\includegraphics[width=0.9\linewidth]{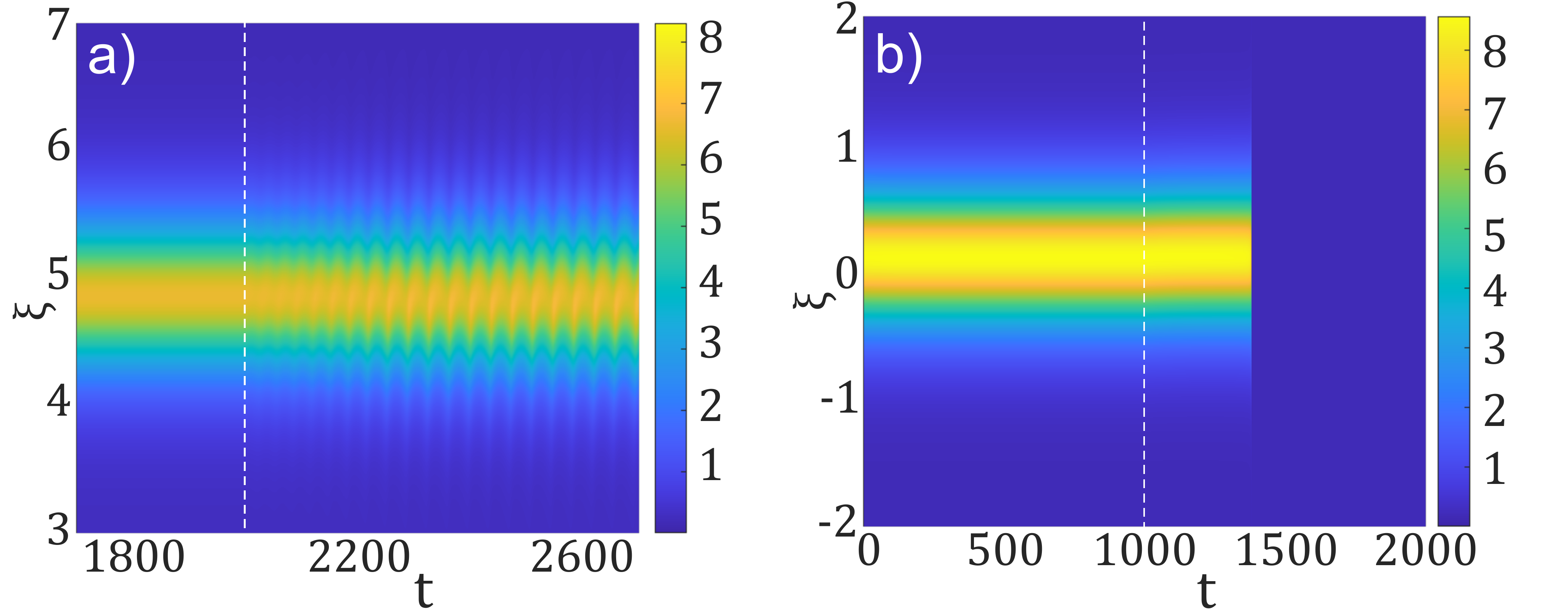}
\caption{Transition from stationary to oscillating TCS via an AH bifurcation for $\theta = 3.5$ (a). $V = 0.053 < V_{AH}$ ($V = 0.0544 > V_{AH}$) for $t<2\times10^3$ ($t>2\times10^3$). TCS collapse after a SN bifurcation for $\theta = 4.427$ (b). $V = 0.0009 < V_{sn}$ ($V = 0.001 > V_{sn}$) for $t<10^3$ ($t>10^3$). Pulse parameters are $p_0 = 1.9$, $d = 0.005$ and $c = 0$. }
\label{fig:desynch}
\end{figure}

If the repetition frequency of the injection pulses differs from the cavity free spectral range, the TCS experiences an additional drift caused by the presence of the derivative term proportional to the parameter $V$ in Eq.~(\ref{eq:PDE_V}). A stationary TCS can only exist if the drift caused by the frequency mismatch is compensated by the opposite drift resulting from the injection gradient. Figure~\ref{fig:synch_reg}(a,b) shows the dependence of the peak intensity and the displacement $\xi_s$ of stationary TCS  excited by Gaussian pulses on the drift parameter $V$ when $p_0 < p_c$. As $V$ increases, the TCS becomes unstable and loses synchronization through an AH bifurcation at the point $V=V_{AH}$.   This bifurcation leads to the TCS oscillating in amplitude and coordinate, which exists in a small parameter range beyond the AH point. 
The transition to an oscillating TCS is illustrated in Fig.~\ref{fig:desynch}(a).  As the drift parameter continues to increase, two unstable TCSs merge at the SN bifurcation point $V=V_{sn}$ and disappear. Another desynchronization scenario is illustrated in Fig.~\ref{fig:desynch}(b), where an AH bifurcation is absent and the TCS remains stable until the SN bifurcation. The corresponding behavior of the eigenvalue spectrum for both scenarios can be seen in Fig.~S1 (Supplement 1).  However, the range of detunings $\theta$ where this second scenario occurs tends to zero in the limit of very broad injection pulses.



The dependence of the TCS peak intensity on the drift parameter $V$ looks different when the injection pulse peak intensity exceeds the pitchfork bifurcation threshold, $p_0>p_c$, see Fig.~\ref{fig:synch_reg}(c). Here, in the absence of frequency mismatch ($V = 0$), the TCS solution located at $\xi_s=0$ [the highest point in Fig.~\ref{fig:synch_reg}(c)] is destabilized by a pitchfork bifurcation, and two stable solutions shifted from $\xi=0$ appear. These solutions correspond to the intersection of the blue and red lines in Fig.~\ref{fig:synch_reg}(c). 
Since two stable TCSs are initially shifted from the origin, the effect of frequency mismatch $V$ and $-V$ on them is asymmetric. With a decrease (increase) of the frequency mismatch, right (left) shifted TCS merges with the destabilized TCS at the SN bifurcation point $V_{sn1}$ ($-V_{sn1}$). As it increases (decreases), such a TCS first loses its stability through an AH bifurcation at the point $V_{AH}$ ($-V_{AH}$) and then merges with an unstable TCS at the SN bifurcation point $V_{sn2}$ ($-V_{sn2}$). As in Fig.~\ref{fig:synch_reg}(b), panel (d) shows the displacement of the TCS from the center of the injection pulse. It can be clearly seen that for both $p_0<p_c$ and $p_0>p_c$ the AH bifurcation occurs when the displacement of the TCS from the pulse center is maximal.

Note that if the frequency mismatch continues to decrease (increase) after reaching the SN point $V_{sn1}$ ($-V_{sn1}$), the solution drops to the nearest stable TCS branch, see Fig.~S2 (Supplement 1). 
Since such transitions between different branches of stable TCSs are possible without loss of synchronization, similar to the case $p_0<p_c$ for $p_0>p_c$ the synchronization range is limited by an AH bifurcation.

Under inhomogeneous pumping, a TCS experiences a drift proportional to the gradient of the injection inhomogeneity. On the other hand, if the repetition rate of the injection pulses slightly deviates from the cavity FSR the TCS also experiences a drift with the velocity $V$, which is proportional to the difference in repetition frequencies. As reported in \cite{hendry2019impact,hendry2020novel,erkintalo2022phase}, when both the inhomogeneity gradient and the drift parameter $V$ are sufficiently small, the slow evolution of the TCS position is governed by the equation:
\begin{equation}
\frac{d\xi_t}{dt}=-V+ \eta_{11}\braket{\Re\psi_{1}^\dagger}{\zeta}+\eta_{12}\braket{\Im\psi_{1}^\dagger}{\zeta},\label{eq:drift}
\end{equation}
where $\xi_t$ is the TCS coordinate and $\zeta=\xi-\xi_t$. The function $\psi_{1}^\dagger=\psi_{1}^\dagger(\zeta)$ is the translational neutral mode of the operator adjoint to the linear operator ${\cal L}_{0}$ (see Supplement 1) describing the TCS stability in the LLE with homogeneous injection $\eta = \eta(\xi_t)$.  
The quantities $\eta_{11}=\Re\left(\partial_\xi\eta\right)_{\xi=\xi_t}$ and $\eta_{12}=\Im\left(\partial_\xi\eta\right)_{\xi=\xi_t}$ define the injection gradient evaluated at $\xi=\xi_t$. In Ref.~\cite{firth1996optical} it was shown that for real $\eta(\xi_t)$ one gets $\braket{\Im\psi_{1}^\dagger}{\xi}\eta(\xi_t) = 2$ in Eq.~(\ref{eq:drift}).

\begin{figure}[ht]
\centering
\includegraphics[width=0.9\linewidth]{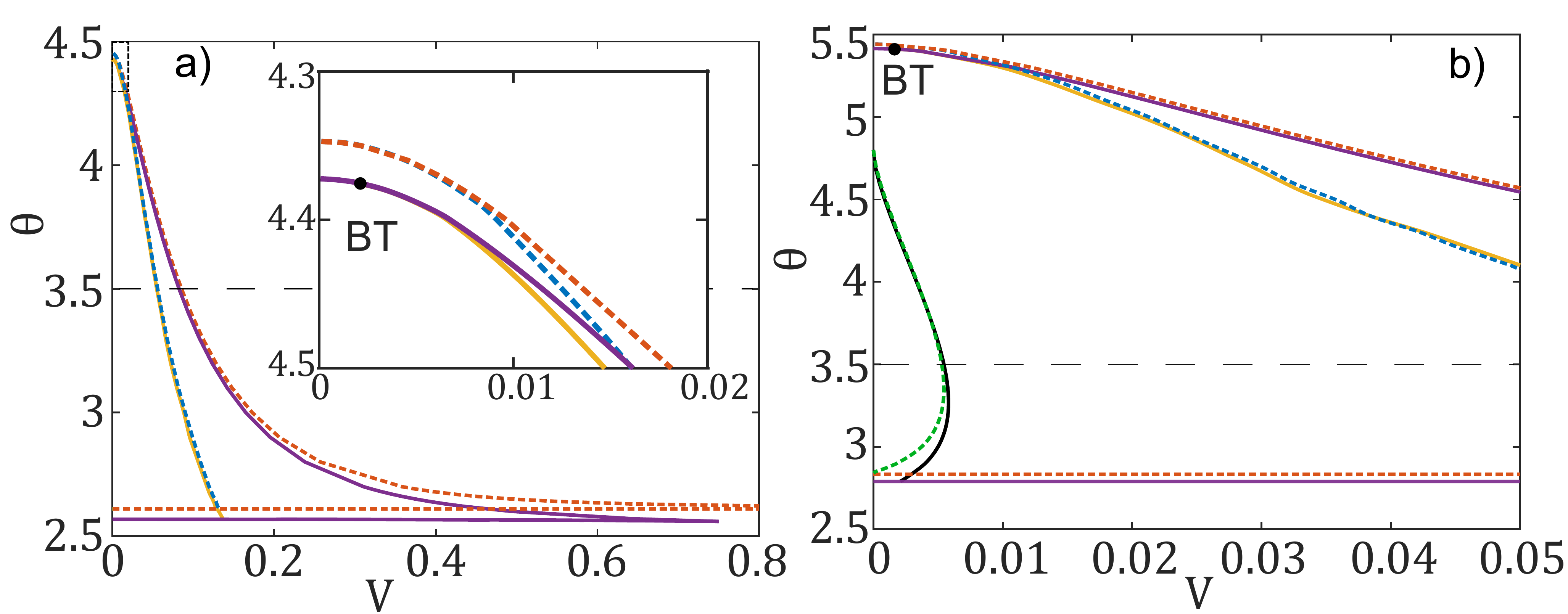}
\caption{Numerical and asymptotic synchronization boundaries for a pulse with $p_0 = 1.9<p_c$, $d = 0.005$, and $c=0$ (a). The inset shows a zoomed-in region at the top of the synchronization domain. Same as (a) but for $p_0 = 2.1>p_c$ (b). Solid (dashed) lines show numerical (asymptotic) bifurcation curves: yellow and blue for AH; red, violet, green and black for SN.}
\label{fig:num_analyt_sync_region}
\end{figure}

According to Eq.~(\ref{eq:drift}) for a stationary TCS with the repetition frequency locked to that of the injection pulses the right-hand side of this equation must be equal to zero. This condition can be used to determine the TCS trapping position $\xi_t=\xi_s$. Without loss of generality, we choose the phase of the field in such a way that $\eta(\xi_s)$ is real.  If $V$ is sufficiently large, larger than the maximum of the sum of the second and third terms in the right-hand side of Eq.~(\ref{eq:drift}) synchronization cannot be achieved and stationary TCS does not exist. The condition that the right-hand side of Eq.~(\ref{eq:drift}) is zero gives an asymptotic estimate of the TCS SN bifurcations at $V=\pm V_{sn}$ shown in Fig.~\ref{fig:synch_reg}(a,b). In Fig.~\ref{fig:num_analyt_sync_region} the curve corresponding to this asymptotic condition 
appears to be very close to the numerical TCS SN bifurcation curve.

\begin{figure}[ht]
\centering
\includegraphics[width=0.9\linewidth]{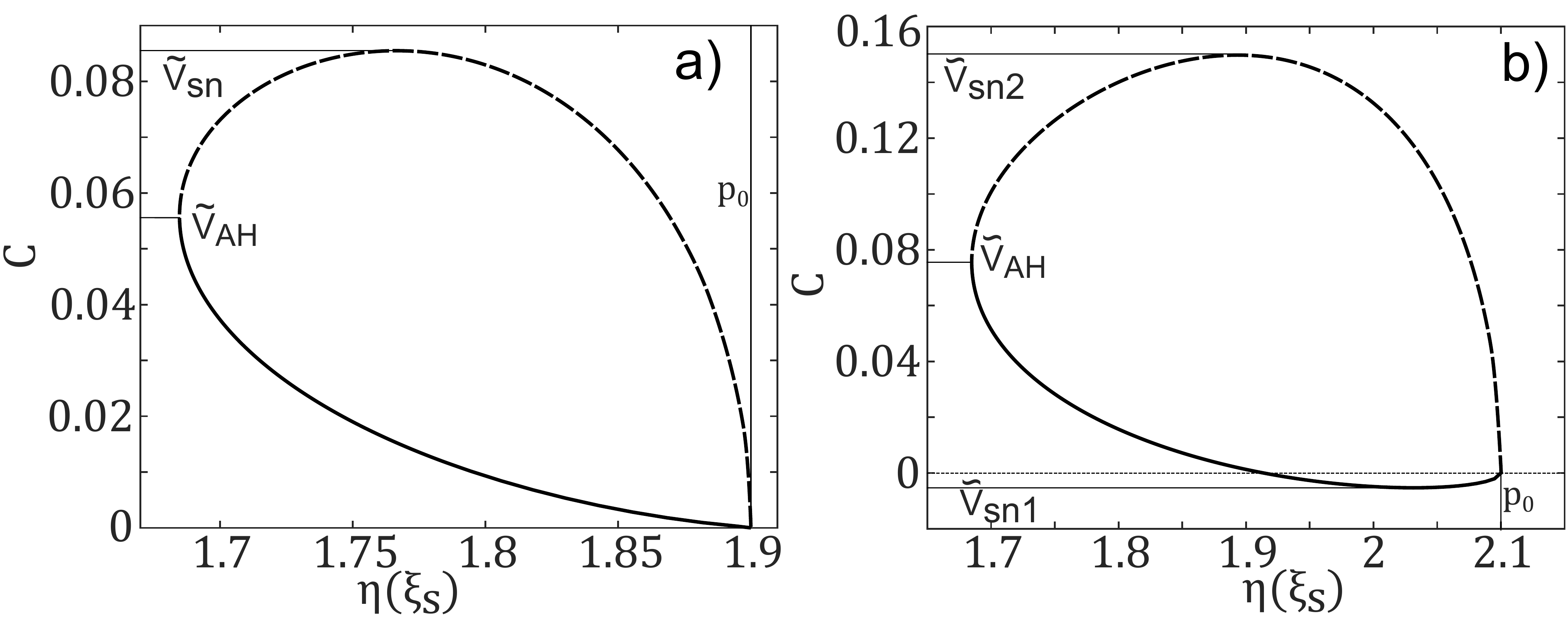}
\caption{Dependence of the quantity $C$ on $\eta(\xi_s)$ for an injection pulse used in Fig.~2(a,b) (left) and Fig.~2(c,d) (right). Solid (dashed) part of the curves corresponds to $C$ calculated with dynamically stable (unstable) TCS under homogeneous injection. 
}
\label{fig:koef}
\end{figure}

The dependence on $\eta(\xi_s)$ of the second term $C = \eta_{11} \braket{\Re\psi_1^{\dagger}}{\xi}$ from the right-hand side of Eq.~(\ref{eq:drift}) for Gaussian pulses from Fig.~\ref{fig:synch_reg} is shown in Fig.~\ref{fig:koef}. Here the last term is zero since the injection pulses are purely real. 
The maximum value of $C=C_{max}$ in Fig.~\ref{fig:koef}(a) equal to $\Tilde{V}_{sn}$ agrees well with the numerically calculated SN point $V_{sn}$ in Fig.~\ref{fig:synch_reg}(a,b). The maximum (minimum) value of $C_{max}$ ($C_{min}$) in Fig.~\ref{fig:koef}(b) equal to $\Tilde{V}_{sn2}$ ($\Tilde{V}_{sn1}$) agrees well with the numerically calculated $V_{sn2}$ ($V_{sn1}$) in Fig.~\ref{fig:synch_reg}(c,d).
Notably, the value of $C$ calculated at the minimum value of $\eta(\xi_s)$ agrees well with the numerically calculated AH bifurcation point, $C(\eta_{min}) = \tilde{V}_{AH} \approx V_{AH}$. Note that  similar rules work for chirped injection pulses when the third term in the right-hand side of Eq.~(\ref{eq:drift}) is nonzero,  
see Fig.~S3 (Supplement 1). 


Although Eq.~(\ref{eq:drift}) provides a good approximation for the TCS SN bifurcation, our simulations show that the desynchronization threshold is typically determined by the AH bifurcation rather than the SN bifurcation. Therefore, an asymptotic analysis of the AH bifurcation boundary is presented below. The frequency mismatch $V$ shifts the stationary TCS towards the edge of the injection pulse. This gradually reduces the injection level until it reaches the critical value of $|\eta|=\eta_0$ at the point $\xi_s=\xi_0$, which corresponds to the TCS SN bifurcation in the LLE with homogeneous injection.
 No stable TCS can exist under homogeneous injection for $|\eta|<\eta_0$.

Let us separate real and imaginary parts of the LLE ~(\ref{eq:PDE_V}). Then we get two real equations for the components of the vector 
 $\vec{A}=\left(\begin{matrix}\Re A & \Im A\end{matrix}\right)^{T}$. 
Near the point $\xi_0$ the injection vector $\vec{\eta}=\left(\begin{matrix}\Re \eta & \Im \eta\end{matrix}\right)^{T}$ can be expanded as ${\vec\eta}\approx\eta_{0}\left(\begin{matrix}1 & 0\end{matrix}\right)^{T}+\epsilon\zeta \vec{\eta}_{1}+\epsilon^{2}\zeta^{2}\vec{\eta}_{2}$ with $\vec{\eta}_{1}=\left(\partial_{\xi}\vec{\eta}\right)_{\xi=\xi_{0}}$, $\vec{\eta}_{2}=\left(\partial_{\xi\xi}\vec{\eta}\right)_{\xi=\xi_{0}}/2$
and $\zeta=\xi-\xi_{0}$. The drift parameter can be written as $V\approx\epsilon v_{0}+\epsilon^{2}v_{1}$.

We look for the solution of the LLE in the form 
\begin{gather}
\vec{A}\left(\zeta,t\right)\approx\vec{A}_{0}[\zeta-\epsilon b(\tau)]+\epsilon a\left(\tau\right)\vec{\psi}_{0}[\zeta-\epsilon b(\tau)\nonumber]\\
+ \epsilon\vec{A}_{1}[\zeta-\epsilon b(\tau)]+\epsilon^{2}\vec{A}_{2}[\zeta,\tau],\label{eq:A}
\end{gather}
where the slow time is $\tau=\epsilon t$ and the neutral mode $\vec{\psi}_{0}(\zeta)=\vec{\psi}_{0}(-\zeta)$  is an even eigenfunction of the linear operator ${\cal L}_{0}$,
describing the stability of the unperturbed solution $\vec{A}_{0}(\zeta)
$ of Eq.~(\ref{eq:PDE_V}) with $V=0$ and constant $\eta=\eta_{0}$. 

Substituting Eq.~(\ref{eq:A}) into the LLE
and collecting the zeroth order terms in $\epsilon$, we obtain the equation for the unperturbed TCS, 
which is automatically fulfilled. Collecting the first order terms in $\epsilon$ and using the relations ${\cal L}_{0}\vec{\psi}_{0,1}=0$, where $\vec{\psi}_{1}(\zeta)=\partial_{\zeta}\vec{A}_{0}=-\vec{\psi}_{1}(-\zeta)$ is the odd translational neutral mode of ${\cal L}_0$, we get the equation

$-{\cal L}_{0}\vec{A}_{1}=\zeta \vec{\eta}_{1}-v_{0}\vec{\psi}_{1}$. The solvability of this equation requires that the right-hand side is orthogonal to $\vec{\psi}_{1}^{\dagger}$. Thus, we recover Eq.~(\ref{eq:drift}) with $\xi_s=\xi_0$ and $\partial_t \xi_0=0$ which can be used to determine the drift parameter $v_{0}=\eta_{11}\braket{ \Re\vec{\psi}_{1}^{\dagger}}{\zeta}+\eta_{12}\braket{ \Im\vec{\psi}_{1}^{\dagger}}{\zeta}$.
Note, that similar to $\vec{\psi}_{1}$, the first order correction $\vec{A}_{1}$ is an odd function of $\zeta$. Then, collecting the second-order terms in $\epsilon$ we get:
\begin{gather}
-{\cal L}_{0}{\vec A}_{2}=-\vec{\psi}_{0}\partial_{\tau}a + \vec{\psi}_{1}\partial_{\tau}b+\zeta^{2}\vec{\eta}_{2}-v_{1}\partial_{\zeta}\vec{A}_{0}-\left(v_{0} + 1\right)\partial_{\zeta}\vec{A}_{1}\nonumber\\-av_{0}\partial_{\zeta}\vec{\psi}_{0} +\vec{\eta}_1 b + {\vec{N}}, \label{eq:second_order}     \end{gather}
where the two components of the vector ${\vec{N}}$ are quadratic forms of $\vec{\Delta A}=a\vec{\psi}_0 + {\vec A}_1$ given in the Supplement 1.

Solvability conditions of this equation give two equations (here we have used the properties of evenness of the functions $\vec{A}_{0}$, $\vec{\psi}_{0}$ and $\zeta^{2}$ and oddness of the functions $\vec{A}_{1}$ and $\vec{\psi}_{1}$):
\begin{gather}
\partial_{\tau}a=g_{0}+g_{1}b+g_{2}a^{2}
\nonumber\\
\partial_{\tau}b=v_{1}+g_{3}a.\label{eq:ab}
\end{gather}
The expressions for $g_{0,1,2,3}$ are given in the Supplement 1.

The characteristic equation of the system (\ref{eq:ab}) reads:
\[
\lambda^2-2ag_2\lambda-g_1g_3=0.
\]
The AH bifurcation occurs at the steady state solution with $a=0$, which corresponds to $v_{1}=0$ in Eqs.~(\ref{eq:ab}). This means that the AH bifurcation occurs
at $V=\epsilon v_{0}$ corresponding to the point $\xi=\xi_{0}$
with $\eta=\eta_{0}$. For the parameters taken from Fig.~\ref{fig:synch_reg}(a,b) AH bifurcation occurs with $\lambda \approx \pm 0.2113 i$ (see the corresponding values of $g_{0,1,2,3}$ in Supplement 1). The eigenvalues of the numerically calculated TCS at the AH point are $\lambda_{num} \approx \pm 0.2059 i$, which is in good agreement with the asymptotic prediction. In our simulations, when the TCS reaches the injection level $\eta=\eta_{0}$, it undergoes an AH
bifurcation. Subsequently, with a further increase in the drift parameter
$V$, the resulting unstable TCS is shifted
back to larger injection levels, $\eta>\eta_{0}$. The asymptotic prediction of $V_{AH}$ agrees well with the numerically performed stability analysis of TCS solutions, see Fig.~\ref{fig:num_analyt_sync_region}. Besides, numerical analysis shows the presence of a Bogdanov-Takens (BT) bifurcation, where the AH bifurcation curve meets the SN one. However, in a limit of the infinitely wide pulse ($\epsilon \rightarrow 0$), BT point shifts to the top of the synchronization region at $V = 0$, see Fig.~S4 (Supplement 1). This is consistent with the fact that the stationary state of Eqs.~(\ref{eq:ab}) has a double zero eigenvalue with geometric multiplicity one when $v_1=0$ and $\vec{\eta}_1=0$ (and, hence, $g_1 = 0$) since zero injection gradient must correspond to zero drift parameter $V$.

In conclusion, we conducted a comprehensive study of the bifurcation mechanisms that cause the TCS repetition rate to desynchronize from that of the injection pulses in a synchronously pumped optical microcavity modeled by the LLE. Both the moderate and large injection peak power cases are considered. As previously demonstrated for moderate power, the TCS remains at the top of the injection pulse at zero repetition rate detuning. However, for large injection peak power, the TCS shifts to the periphery of the injection pulse \cite{hendry2018spontaneous,hendry2019impact,hendry2020novel,erkintalo2022phase,talenti2023control}. Our study shows that desynchronization usually occurs through an AH bifurcation, which limits the locking range of the TCS when the injection pulse is wide enough. We have presented a straightforward and easy-to-understand criterion for identifying this bifurcation. It occurs when the TCS shift reaches a maximum and coincides with the point at which the amplitude of the injection pulse decreases to a level equivalent to the TCS SN bifurcation observed in LLE with homogeneous injection $\eta=\eta_0$.Therefore, to determine the soliton displacement $\xi_0$ at the desynchronization threshold, it is necessary to know only the injection pulse profile and $\eta_0$. The critical mismatch $V$ can then be determined from Eq.~(\ref{eq:drift}). Numerical and asymptotic evidence is presented to substantiate the reliability of this criterion.

The support by the Deutsche Forschungsgemeinschaft
(DFG project No. 491234846) is
gratefully acknowledged.







\noindent{\bf Disclosures.} The authors declare no conflicts of interest.

\noindent{\bf Data availability.} Data underlying the results presented in this paper are available upon reasonable request.

\noindent{\bf Supplemental document.} See Supplement 1 for supporting content.

\bibliography{comb}

\bibliographyfullrefs{comb}

\end{document}


\maketitle

\section{Asymptotic analysis}

The linear operator ${\cal L}_{0}$ used in the asymptotic analysis is given by:
\[
{\cal L}_{0}=\left(\begin{array}{cc}
-1-2X_{0}Y_{0} & -\partial_{\zeta\zeta}+\theta-X_{0}^{2}-3Y_{0}^{2}\\
\partial_{\zeta\zeta}-\theta+3X_{0}^{2}+Y_{0}^{2} & -1+2X_{0}Y_{0}
\end{array}\right),
\]
where $X_0 = \Re A_0$ and $Y_0 = \Im A_0$ are two components of the vector  $\vec{A_0}(\zeta)=\left(\begin{matrix}X_0 & Y_0\end{matrix}\right)^{T}$  of the unperturbed temporal cavity soliton (TCS) solution  obtained under homogeneous injection $\eta(\zeta) = \eta_0$.

The last term $\vec{N}$ in the right-hand-side of Eq.~(4) can be written in the form :
\[\vec{N} = \dfrac{1}{2}\left(\begin{array}{c} \vec{\Delta A}^T {\cal H}_1 \vec{\Delta A} \\ 
\vec{\Delta A}^T {\cal H}_2 \vec{\Delta A}\end{array}\right),\]
where $\vec{\Delta A} = a\vec{\psi}_{0} + \vec{A}_{1}$ and ${\cal H}_{1,2}$ are Hessian $2\times2$ matrices of second-order partial derivatives for two equations obtained by separating the real and imaginary parts of The Lugiato-Lefever equation and are given by the following matrices:
\[
{\cal H}_{1} = \left(\begin{array}{cc}
-2Y_0 & -2X_0 \\
-2X_0 & -6Y_0
\end{array}\right), \qquad
{\cal H}_{2} = \left(\begin{array}{cc}
6X_0 & 2Y_0 \\
2Y_0 & 2X_0
\end{array}\right).
\]
Alternatively, this term can be written in the form  $\vec{N} = {\cal N}_{0} \vec{Q}$, where ${\cal N}_{0}$ is a $2\times3$ matrix:
\[
{\cal N}_{0}= \dfrac{1}{2}\left(\begin{array}{ccc}
\partial_{X_0}{{\cal L}_{0}^{1,1}} & \partial_{Y_0}{{\cal L}_{0}^{1,1}} + \partial_{X_0}{{\cal L}_{0}^{1,2}} & \partial_{Y_0}{{\cal L}_{0}^{1,2}}\\
\partial_{X_0}{{\cal L}_{0}^{2,1}} & \partial_{Y_0}{{\cal L}_{0}^{2,1}} + \partial_{X_0}{{\cal L}_{0}^{2,2}} & \partial_{Y_0}{{\cal L}_{0}^{2,2}}
\end{array}\right) = \left(\begin{array}{ccc}
-Y_{0} & -2X_{0} & -3Y_{0}\\
3X_{0} & 2Y_{0} & X_{0}
\end{array}\right)\]
and 
\[\quad\vec{Q}=\left(\begin{array}{c}
{\vec {\Delta A}}_{1}^{2}\\
{\vec {\Delta A}}_{1}{\vec {\Delta A}}_{2}\\
{\vec {\Delta A}}_{2}^{2}
\end{array}\right)
\]
with ${\vec {\Delta A}}_{k}=a\psi_{0k} + A_{1k}$. The index $k=1$ ($k=2$) denotes the real (imaginary) component of the corresponding vector.


The expressions for the coefficients in Eq.~(5) are  

\[g_0 = \braket{\vec{\psi}_{0}^{\dagger}}{\zeta^2 \vec{\eta}_2} - \braket{\vec{\psi}_{0}^{\dagger}}{\partial_{\zeta} \vec{A}_1} \left(v_0 + 1\right) + \braket{\vec{\psi}_{0}^{\dagger}}{{\cal N}_{0} \vec{n}_0},
\]

\[g_1 = \braket{\vec{\psi}_{0}^{\dagger}}{\vec{\eta_1}},
\]

\[g_2 = \braket{\vec{\psi}_{0}^{\dagger}}{{\cal N}_{0} \vec{n}_1},
\]

\[g_3 = \braket{\vec{\psi}_{1}^{\dagger}}{\partial_{\zeta} \vec{\psi}_0} v_0 - \braket{\vec{\psi}_{1}^{\dagger}}{{\cal N}_{0} \vec{n}_2},
\]
where \[\vec{n}_0 = \left(\begin{array}{c}
A_{11}^{2}\\
A_{11}A_{12}\\
A_{12}^{2}
\end{array}\right),  \quad
\vec{n}_1 = \left(\begin{array}{c}
\psi_{01}^2\\
\psi_{01}\psi_{02}\\
\psi_{02}^2
\end{array}\right),  \quad
\vec{n}_2 = \left(\begin{array}{c}
2 A_{11} \psi_{01}\\
A_{12}\psi_{01} + A_{11}\psi_{02}\\
2 A_{12} \psi_{02}
\end{array}\right)\]
and the neutral modes satisfy the biorthogonality conditions $\braket{\vec{\psi}_j^{\dagger}}{\vec{\psi}_k} =\int_{-L/2}^{L/2} \left(\psi_{j1}^{\dagger}\psi_{k1}+\psi_{j2}^{\dagger}\psi_{k2}\right)d\xi= \delta_{j,k}$ with  $k,j=0,1$.

\textbf{Numerical example.} For the injection pulse with $d = 0.005$, $p_0 = 1.9$, and $\theta = 3.5$, which corresponds to the parameters of Eq.~(1) from Fig.~2(a,b), the values of numerically calculated coefficients are: 
\[g_0 = 1.7347,
g_1 = -2.9365,
g_2 = -0.36937,
g_3 = 0.015204.
\]

\section{Numerical analysis}

\begin{figure}[hb]
\centering
\includegraphics[width=\linewidth]{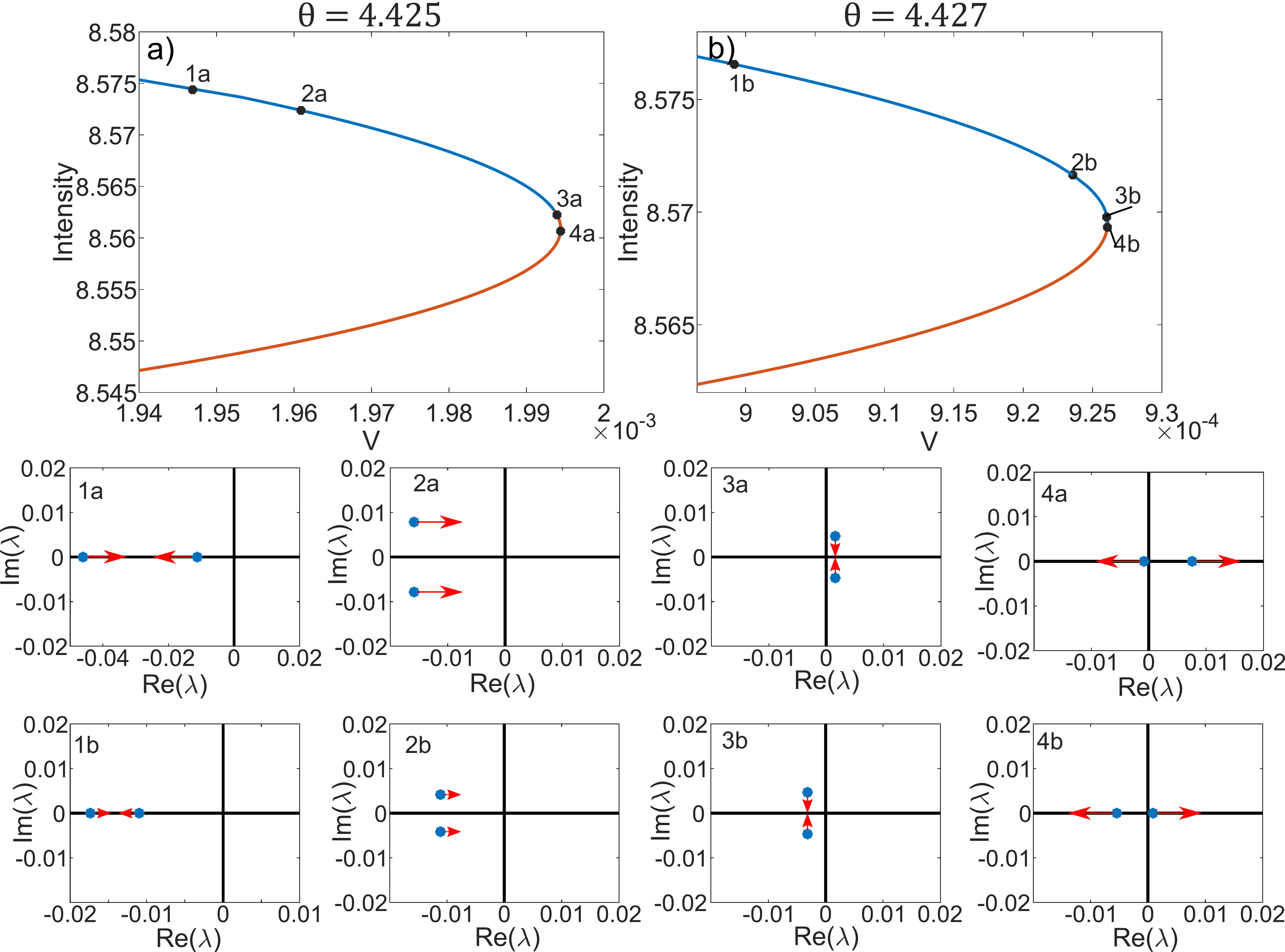}
\caption{Two types of desynchronization transition. 
TCS is destabilized by AH bifurcation before SN one (a).  TCS is destabilized by SN bifurcation (b). Panels 1a-4a show the behavior of two discrete TCS eigenvalues on the complex plane responsible for the desynchronization scenario shown in panel (a). Panels 1b-4b show the behavior of two discrete TCS eigenvalues responsible for the desynchronization scenario shown in panel (b). In (a,b) the blue (red) curve shows stable (unstable) solutions. Parameters are: $p_0 = 1.9$, $d = 0.005$, $c = 0$.}
\label{fig:eig_vals}
\end{figure}

The behavior of the eigenvalues illustrating two possible scenarios of TCS desynchronization with increasing $V$ is shown in Fig.~\ref{fig:eig_vals}. The first scenario shown in panel (a), where the TCS is first destabilized by an Andronov-Hopf (AH) bifurcation, is as follows: two negative real eigenvalues collide (panel 1a) and form a complex conjugate pair. Then this pair approaches the imaginary axis (panel 2a) and crosses it, causing an AH bifurcation. After that, the complex pair collides again in the right half of the complex plane (panel 3a) and forms two real eigenvalues (panel 4a), one of which crosses zero and provides the saddle-node (SN) bifurcation. In the second scenario shown in Fig.~\ref{fig:eig_vals}(b), there is no AH bifurcation and the second collision of the eigenvalues takes place in the left half of the complex plane (panel 3b), eventually leading to a SN bifurcation. A case where the second collision takes place on the imaginary axis corresponds to a Bogdanov-Takens bifurcation. It is important to note that both eigenvalues shown in Fig.~\ref{fig:eig_vals} have their counterparts in the case of homogeneous injection. The eigenvalue closest to zero [in panels (1a) and (1b)] corresponds to the zero eigenvalue associated with the translational neutral mode in the homogeneous case.  The second eigenvalue is responsible for the SN bifurcation of the TCSs in Eq.~(1) with homogeneous injection. Thus, in the homogeneous case, two eigenvalues correspond to the neutral modes ${\vec \psi}_{0,1}$ with different evenness.

\begin{figure}[ht]
\centering
\includegraphics[width=\linewidth]{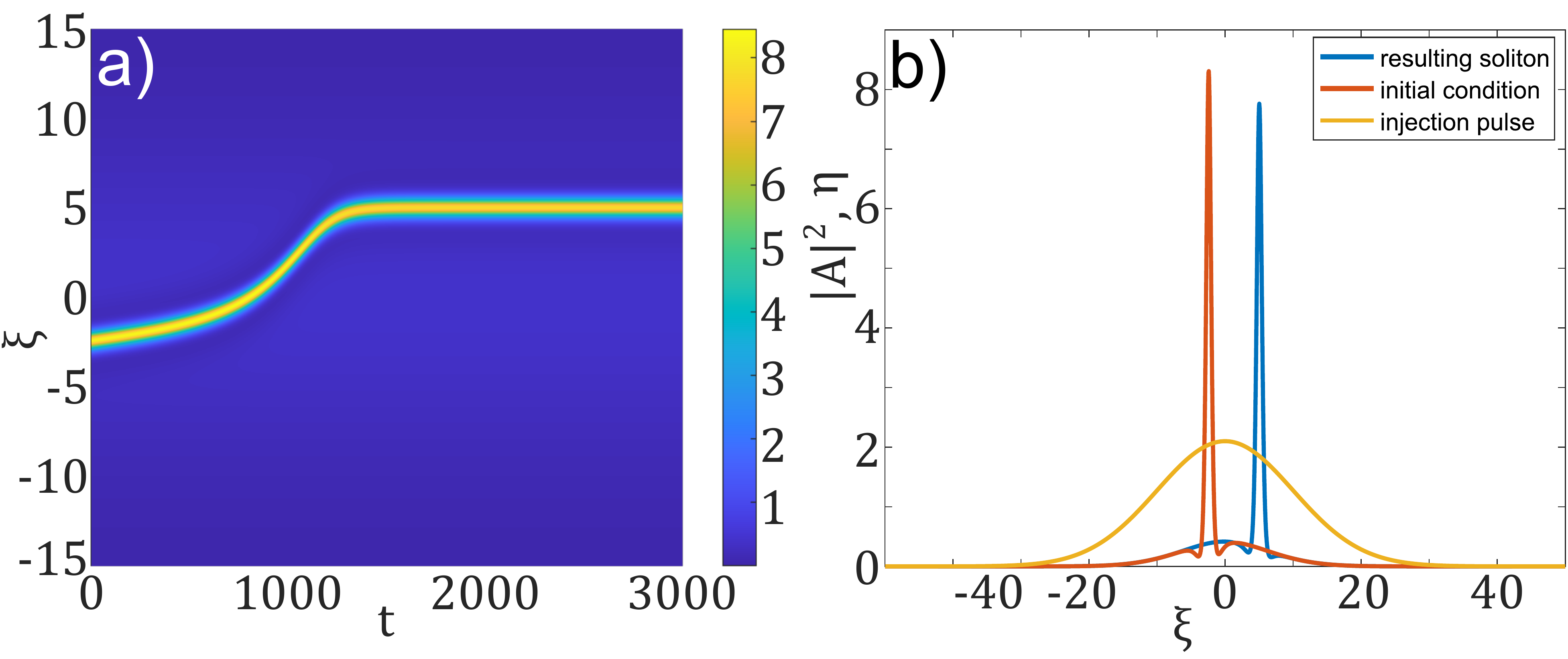}
\caption{Transition between different parts of the TCS branch obtained by numerical integration of Eq.~(1) with parameters taken from Fig.~2(c,d) ($p_0 = 2.1$, $d = 0.005$, $c = 0$) (a). This transition is indicated by arrow in  Fig.~2(d). The value $V$ is slightly larger than $V_{sn1}$ ($V = 0.007$) and an initial condition corresponds to the shifted TCS at $V = V_{sn1}$.  
Panel b) shows the profiles of the initial condition (red), the resulting TCS (blue), and the injection pulse (yellow).}
\label{fig:jump_sn}
\end{figure}

The transition between two branches of the TCS with sufficiently large $p_0 > p_c$ is illustrated in Fig.~\ref{fig:jump_sn}. This figure shows the results of numerical integration of Eq.~(1) with $V > V_{sn1}$ and the TCS corresponding to $V_{sn1}$ as initial condition. In panel (a) one can see how the TCS, initially shifted to the left of the injection pulse peak, passes through the pulse center and ends up at the right tail of the pulse. Panel (b) shows the initial and final field distributions and the shape of the injection pulse. Such a transition corresponds to a jump from the SN point with $V=V_{sn1}$ to the other branch of stable TCSs, see the arrow in Fig.~2(c) in the main text.
\begin{figure}[ht]
\centering
\includegraphics[width=\linewidth]{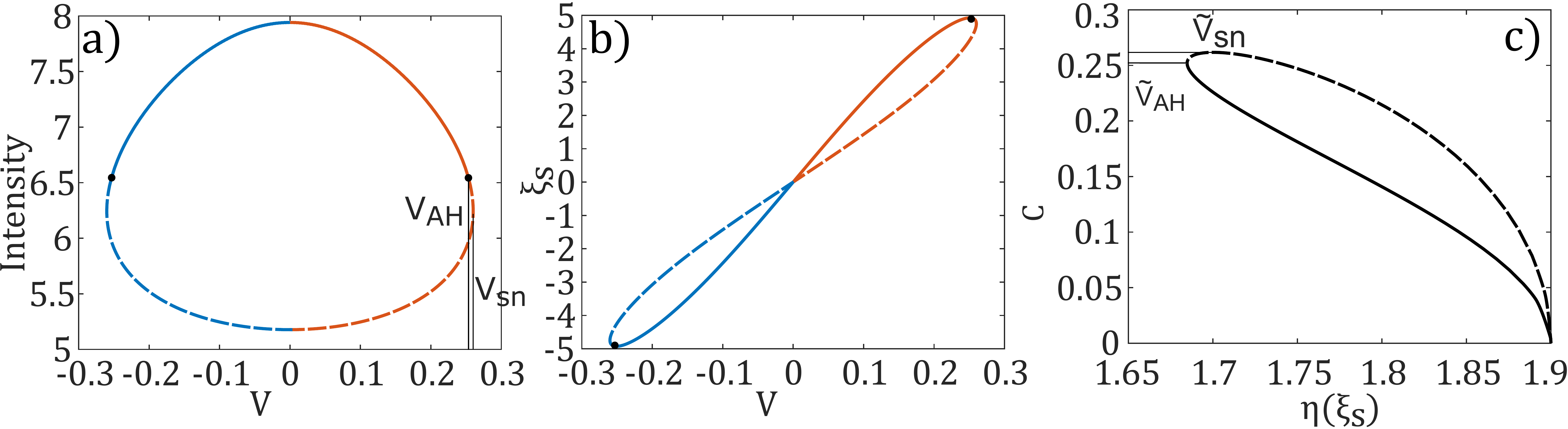}
\caption{Same as Fig.~2(a,~b), but with chirped injection pulse  ($c = 0.01$) (a,b). Dependence of  the quantity $C$ on the local pump $\eta(\xi_s)$ for the chirped pulse (c).} 
\label{fig:chirp}
\end{figure}

Synchronization by a chirped injection pulse is illustrated in Fig.~\ref{fig:chirp}. 
In this case, the last term on the right-hand side of Eq.~(2) is nonzero and, together with the term proportional to $\eta_{11}$, contributes to the compensation of the drift $V$. Comparing Fig.~\ref{fig:chirp}(a) and Fig.~2(a), where the same form of a pulse is used without chirp, it can be seen that the synchronization range increases with the chosen form of the chirp function ($\phi = -c\xi^2$). It is obvious that the opposite sign of the coefficient $c$ would decrease the synchronization range. Panel (b) shows the dependence of the TCS stationary position on $V$. Comparing this panel with Fig.~2(b) obtained in the absence of chirp and remembering that the chirp does not affect the injection pulse amplitude, one can see that the values of the maximum dislocations are close. Therefore we can conclude that the AH bifurcation occurs at a similar injection value $\eta(\xi_s)$. Panel (c) shows the dependence of the sum of the last two terms in Eq.~(2) $C = \eta_{11} \braket{\psi_{11}^{\dagger}}{\xi} + \eta_{12} \braket{\psi_{12}^{\dagger}}{\xi}$ on $\eta(\xi_s)$.

\begin{figure}[h]
\centering
\includegraphics[width=0.5\linewidth]{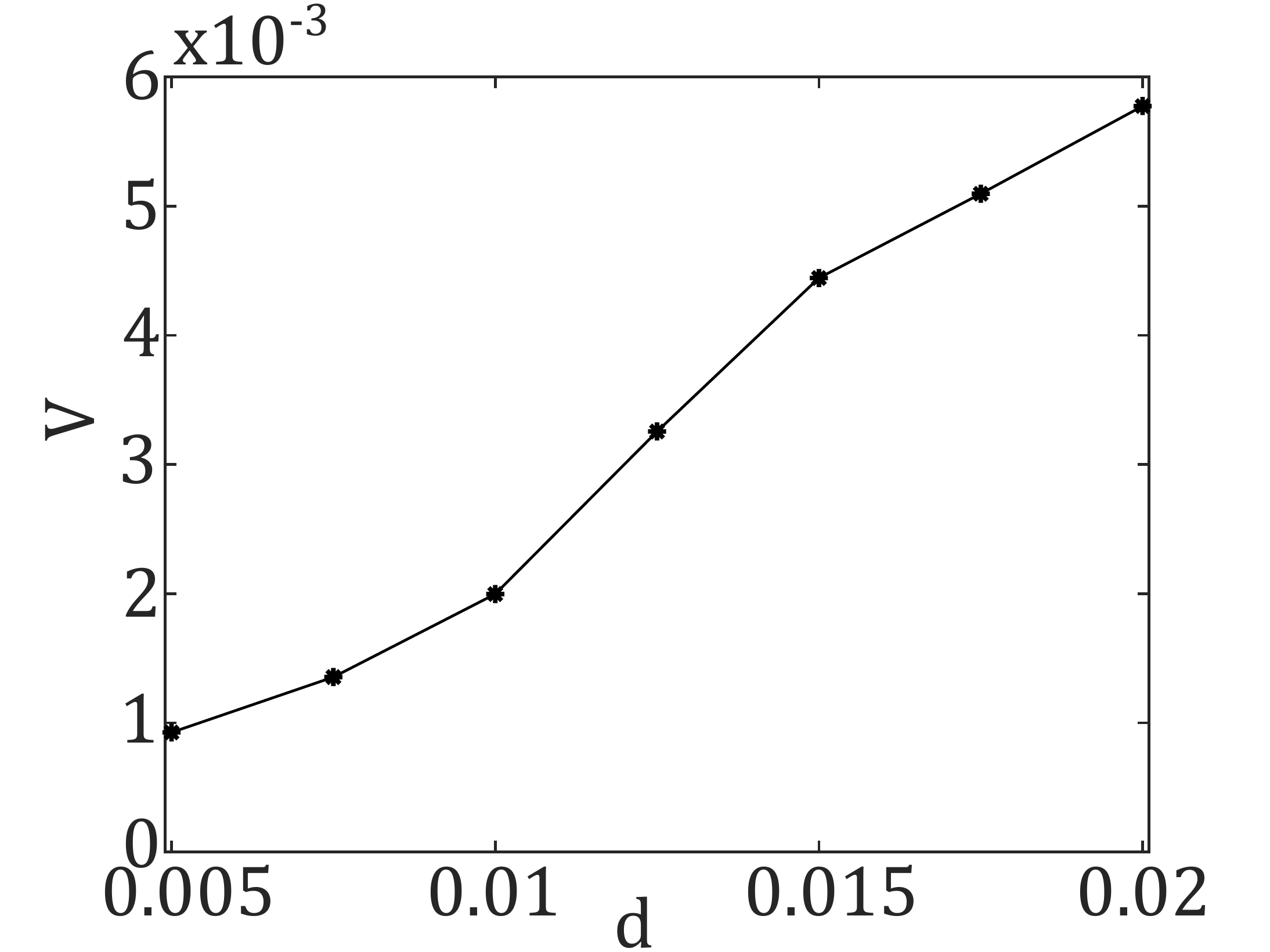}
\caption{Bogdanov-Takens bifurcation points for seven different values of the injection pulse width: $d = 0.005$, $d = 0.0075$, $d = 0.01$, $d = 0.0125$, $d = 0.015$, $d = 0.0175$, $d = 0.02$. Calculations performed for $p_0 = 1.9$.}
\label{fig:BT}
\end{figure}

Fig.~\ref{fig:BT} shows the dependence of the numerically calculated position of the Bogdanov-Takens bifurcation on the width of the injection pulse. The value $d = 0$ corresponds to the homogeneous case and the width of the pulse is inversely proportional to the value $d$. In the limit $d \rightarrow 0$ the Bogdanov-Takens bifurcation point moves to $V = 0$ and $\theta = \theta_{max}$, where $\theta_{max}$ corresponds to the maximum value of detuning above which no soliton exists.
